\documentclass{article}

\usepackage{arxiv}

\usepackage[utf8]{inputenc} 
\usepackage[T1]{fontenc}    
\usepackage{hyperref}       
\usepackage{url}            
\usepackage{booktabs}       
\usepackage{amsfonts}       
\usepackage{nicefrac}       
\usepackage{microtype}      
\usepackage{lipsum}		
\usepackage{graphicx}
\usepackage{doi}

\usepackage[table]{xcolor}

\usepackage[natbibapa,nodoi]{apacite}
\usepackage{multibib} 
\newcites{PS}{Primary Studies}

\usepackage{hyperref}
\usepackage{pgfplots}
\pgfplotsset{compat=1.9}
\usepackage{pgf-pie}
\usepackage{url}

\usepackage{float}
\usepackage{subcaption}
\usepackage{multirow}

\definecolor{formalshade}{rgb}{0.95,0.95,1}

\usepackage{tikz}
\usepackage{calc}
\def\scalecheck{\resizebox{\widthof{\checkmark}*\ratio{\widthof{x}}{\widthof{\normalsize x}}}{!}{\checkmark}} 

\usepackage{hyperref}
\let\svthefootnote\thefootnote
\newcommand\freefootnote[1]{%
  \let\thefootnote\relax%
  \footnotetext{#1}%
  \let\thefootnote\svthefootnote%
}

\title{Towards User-Centric Guidelines for Chatbot Conversational Design}

\date{} 					

\author{ \href{https://orcid.org/0000-0002-0304-0804}{\includegraphics[scale=0.06]{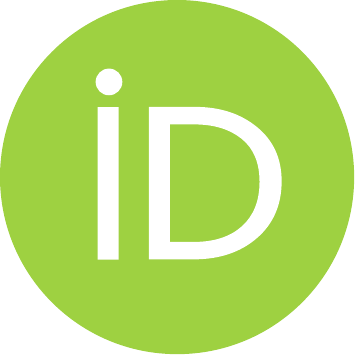}\hspace{1mm}Geovana Ramos Sousa Silva}\\
	Department of Computer Science\\
	University of Brasília (UnB)\\
	Bras\'{i}lia, Brazil \\
    \hfill \\
	\And
	\href{https://orcid.org/0000-0002-2159-339X}{\includegraphics[scale=0.06]{orcid.pdf}\hspace{1mm}Edna Dias Canedo} \\
	Department of Computer Science\\
	University of Brasília (UnB)\\
	Bras\'{i}lia, Brazil \\
 \hfill \\
}

\renewcommand{\shorttitle}{Towards User-Centric Guidelines for Chatbot Conversational Design}

\hypersetup{
pdftitle={Towards User-Centric Guidelines for Chatbot Conversational Design},
pdfsubject={pre-print paper},
pdfauthor={Geovana R. S. Silva, Edna D. Canedo},
pdfkeywords={Chatbot; conversational agents; conversational design; human-AI interaction; conversational user experience},
}

\begin{document}

\maketitle

\begin{abstract}
The conversational nature of chatbots poses challenges to designers since their development is different from other software and requires investigating new practices in the context of human-AI interaction and their impact on user experience. Therefore, this work aims to unveil chatbot conversational practices alongside their impacts on users to build a web guide to support designers while conceiving conversations for chatbots. We have carried out a Systematic Literature Review (SLR) to identify linguist, visual, and interactive elements used in chatbot conversational design. The SLR resulted in 40 selected studies that were reviewed and coded into a set of conversational guidelines that were evaluated through a survey. Respondents strongly agreed that applying the proposed guidelines in chatbot development would induce greater user satisfaction and user engagement and the guide is usable, flexible, clear, and understandable, making it a great ally in building chatbots with an improved user experience.
\end{abstract}


\freefootnote{This is an original manuscript of an article published by Taylor \& Francis in the International Journal of Human Computer Interaction, on 09 Sep 2022, available at: https://doi.org/10.1080/10447318.2022.2118244. It is deposited under the terms of the Creative Commons Attribution-NonCommercial-NoDerivatives License (http://creativecommons.org/licenses/by-nc-nd/4.0/), which permits non-commercial re-use, distribution, and reproduction in any medium, provided the original work is properly cited, and is not altered, transformed, or built upon in any way.} 

\section{Introduction}

The concept of a machine simulating human dialog has been around for a long time, which resulted in a set of different conversational systems. One of these systems is the chatbot, which attempts to simulate human conversations \citep{2020McTear}. In 1966, Joseph Weizenbaum created a conversational robot that simulated a virtual psychotherapist, Eliza, which is considered the first chatbot in history \citep{Weizenbaum66}. 

Since then, chatbots are a trend with a strong presence in business as they streamline and provide uninterrupted services. This trend is also a result of more powerful machine learning algorithms that are capable of building more human-like chatbots, this way companies are more confident to replace the human workforce in customer service. Still, apart from the technological aspect, inadequate conversational decisions from designers can negatively impact users' perception of the chatbot \citep{Zamora17}.

Therefore, chatbot teams are seeing the conversations as an object of design to provide a better user experience \citep{FolstadB17}. Hence, it is necessary to reverse the development process by starting from conversational design and then going to software development. Although there are a lot of available tools for building chatbots such as Rasa and DialogFlow \citep{singh2019introduction}, from the user perspective, there is not a noticeable difference between chatbots with the same conversational flow that are powered by different tools. 

Even though chatbots' capacity to connect to their users evolved, it is still a challenge to mimic human behavior, and user interaction is one of the biggest challenges developers face in chatbot development \citep{Abdellatif2020}. It is important to work on the user experience of talking to a chatbot because it directly impacts the relationship between the user and the organization the chatbot represents. A chatbot with an optimal user experience is capable of communicating accordingly considering its target audience, which consequently makes users more satisfied with the service. Also, following good conversational practices makes users trust the brand behind the chatbot \citep{FolstadNB18}.

Having all that in mind, it is necessary to investigate chatbot conversational requirements that are technology-independent, user-centric, and focused on achieving business goals through conversations. By defining these requirements, it is possible to provide a nicer user experience and positively impact user satisfaction regardless of the technology behind the chatbot. 

Therefore, in this work, we propose guidelines for chatbot design based on a systematic literature review focused on conversational practices and their impacts on users. These guidelines were shared with practitioners to gather their impressions about it and evaluate its usefulness and ease of use. The main findings of our work are a validated set of practices that can be used to enhance human-chatbot interaction and the indication of practices that should be avoided based on their side effects on users. 

\section{Background} 

In this section, we present an overview of chatbots, introduce the concept of conversational design and discuss the similarity and differences of this work with others that approach human-chatbot interaction.

\subsection{Chatbots}

The use of natural language either by text or voice has been widely used in a variety of systems to facilitate and humanize the user interaction with systems \citep{ShneidermanPlaisantEtAl16}. Nowadays, for example, it is possible to change GPS routes without taking your hands off the wheel, book a flight or hotel room by chatting with a machine, search the internet just by talking to your phone or have a virtual agent notify you about important things just like a human would do. In this way, numerous conversational technologies have emerged and they can be categorized in different ways.    

A chatbot is a type of conversational technology that has the following defining features: understanding natural language input, the ability to interact and hold a conversation \citep{cahn2017chatbot}. Still, they can differ in implementation by receiving only text input, voice input, or both. Moreover, their main functions range from performing tasks, troubleshooting, solving doubts, and personal assistance \citep{Adamopoulou2020}. They are strongly present in domains such as education, health, tourism, and general customer service \citep{ADAMOPOULOU2020100006}.

The interest in chatbots rapidly increased in 2016 according to ACM, Google Trends, \citep{marcondes2018chatbot} and Scopus \citep{Adamopoulou2020}. The first thing that boosted this increase that year was the opening for the inclusion of third-party chatbots in major social platforms, such as Slack, Telegram, and Facebook Messenger \citep{yeung_2016}, which would support companies to integrate their own chatbots into their social networks. Having chatbots in the front line of customer service is in the interest of any company as it can replace a team of human attendants and save resources \citep{AndradeSJSMJ20}.

In the following years, the creation of tools for chatbot development that encapsulate the complexity of developing chatbots further stimulated interest in the subject, such as Google's DialogFlow, Microsoft Bot Framework, and Rasa \citep{singh2019introduction}. If in the past developers needed to generate complex machine learning algorithms and build a chatbot from scratch, these tools only require that developers feed the knowledge base and pass the training parameters. 

\subsection{Conversational Design} 

Since the technological barrier and complexity is becoming less of a problem in chatbot development, there has been an increasing interest in human-chatbot interaction to overcome users' resistance in accepting chatbots in comparison with human agents \citep{Lei20213977}. The naturalness of the interaction is a critical aspect to overcome negative preconceptions users may have towards chatbots \citep{Paikari2018}.

Designing chatbot conversations requires research and knowledge of this kind of interaction since users "adapt their language to communicate with intelligent agents" \citep{HILL2015245}. In light of these challenges, conversational design has been in the spotlight in late years. It is an essential process for developing effective chatbots. According to \citet{moore2019}, it comprises the following activities: observe and engage with users, define user personas and goals, shape conversations, define an agent persona, presume user's and agent's messages, prototype and test. 

From a more practical side, according to \citet{mctear2018conversation}, conversation design ensures the promotion of engagement, retention, pleasant customer experience, and measuring the quality of the interaction. It is a multidisciplinary activity that requires the involvement of developers, designers, writers, and business strategists. Designers define the form of interaction and the conversational flow, writers polish communication, developers ensure that the chat platforms support the interactive elements suggested by designers, and business strategists guarantee that the agent represents the brand accordingly.

\subsection{Related Works}\label{sec:related_works}

In late years, chatbot human interaction research is at its high, due to the advance in Artificial Intelligence and consequently the improvement in chatbot interaction capacity. Therefore, since achieving some complexity in conversational design is no longer hidden by the technology available, many studies focus solely on the user experience aspect of human-chatbot interaction.

\citet{chaves2021} conducted a literature review on disembodied, text-based chatbots to derive a conceptual map of social characteristics for chatbots. They analyzed 56 papers and highlighted how social characteristics can beneﬁt human-chatbot interactions, the challenges and strategies to designing them, and how the characteristics may inﬂuence one another. 

The work of \citet{sugisaki2020} discussed how text-based conversational user interfaces (CUIs) are different from other forms of human-computer interaction, and what challenges and opportunities arise from these differences. As a result, they provide a set of 53 technology-agnostic checkpoints specifically for text-based CUIs, which were evaluated by 15 practitioners and academics to examine content validity.

\citet{rapp2021} carried out a systematic literature review of 83 papers that focus on how users interact with text-based chatbots in terms of satisfaction, engagement, and trust, whether and why they accept and use this technology, how they are emotionally involved, what kinds of downsides can be observed in human-chatbot conversations, and how the chatbot is perceived in terms of its humanness.

\citet{ amershi2019guidelines} proposed generally applicable design guidelines for human-AI interaction. The guidelines were first originated from the literature and industry. Then, these guidelines were refined by a modified heuristic evaluation and tested by 49 design practitioners against 20 popular AI-infused products. After revisions, the guidelines were inspected by experts, which resulted in the final set of 18 guidelines.

In the research conducted by \citet{yang2021}, ten guidelines were then derived from an interview study to explore how users’ needs of competence, autonomy, and relatedness could be supported or undermined in experiences with voice assistants. The guidelines recommend how to inform users about the system's capabilities, design effective and socially appropriate conversations, and support increased system intelligence, customization, and data transparency. The interviews also unveiled determining factors for the success of the interaction such as the users' knowledge of the conversational agent capabilities, conversation flexibility, and control over user data. 

\citet{FEINE2019138} conducted a systematic literature review to identify a set of social cues of conversational agents (CA) to develop a taxonomy that classifies the identified social cues into four major categories (i.e., verbal, visual, auditory, invisible) and ten subcategories. The taxonomy was used systematically to identify a wide variety of different social cues implemented in the text-based CA Poncho, in the voice-based CA Alexa, and in the embodied CA SARA.

Although all of these related works are of immense importance for the chatbot community and complement each other, they differ from the present work in focus and in methodology regarding the measures and steps taken in the literature review. \citet{ chaves2021},  \citet{ sugisaki2020}, and \citet{rapp2021} are especially focused on subjective characteristics of human-chatbot interaction while we are more focused on the practical, functional, and interface aspects of conversational design. For example, while they define responsiveness and productivity as important characteristics, we are interested in how they are being achieved and practically implemented in the literature. 

\citet{amershi2019guidelines} had the same focus as the present work, that is, conversational design practices, however, the research refers to the full range of AI products and not only text-based chatbots. Similarly, \citet{FEINE2019138} consider embodied and voice-based CAs whereas we do not. Finally, the work of \citet{yang2021} is the closest to the present work regarding their main objective, however, they define conversational practices through interviews, whereas we are defining through a systematic literature review, which is explained in Section \ref{SLreview}. Although \citet{chaves2021}, \citet{sugisaki2020}, and \citet{rapp2021} are based on literature reviews, besides the differences in their main goal, their review protocols are different from ours.

\section{Systematic Literature Review}
\label{SLreview}

We carried out a Systematic Literature Review (SLR) following the protocol of \citet{ barbara2007guidelines} to unveil chatbot conversational design practices and their impacts on users. The SLR is the process of identifying, evaluating, and interpreting relevant studies of an area or research question of interest \citep{kitchenham2004procedures}, and it is composed of the following phases \citep{barbara2007guidelines}: 

\begin{enumerate}
    \item \textit{Planning}: identifying the need for a review, establishing objectives, and defining the review protocol, which consists of the following artifacts: research questions, search string, study selection criteria, list of data to be extracted, and quality assessment checklist;
    
    \item \textit{Conducting}: it consists of putting the review protocol into practice by utilizing the artifacts produced in the previous phase to filter studies and extract information from them;
    
    \item \textit{Reporting}: documenting the results of the review, in this case, as a research paper.
\end{enumerate}

\subsection{Research Questions}
 
In this work, the motivation for conducting the SLR is to disclose state-of-art practices in chatbot conversational design and how they impact users. It is important to notice that our focus is entirely on conversational design and practices that are directly seen by users, therefore, we are not interested in technical aspects of chatbot development. The research questions are shown in Table \ref{tab:research_questions}.

\begin{table*}[htb!]
  \centering
  \caption{Research Questions}
  \label{tab:research_questions}
  \begin{tabular}{cp{13cm}}
    \hline
    ID & Research Question \\
    \hline
    RQ.1 & What are the textual or visual approaches used in chatbot conversational design?  \\
    RQ.2 & What are the positive or negative impacts on users of the identified textual or visual approaches in chatbot conversational design?  \\
    RQ.3 & Are there moderating effects of other variables on the identified impacts of practices? \\
  \hline
\end{tabular}
\end{table*}

\subsection{Search String}
 
The search string was created through the PICOC method (Population, Intervention, Comparison, Outcome, Context). The population refers to the object of study; the intervention is a procedure executed by or applied to the population; the comparison is what is being compared with the intervention; the outcome is a result of the intervention; and the context is the focus of the study, its restrictions, and limitations. Table \ref{tab:picoc} shows the final definition of the PICOC terms to build the generic search string. Comparison is not applicable because we are not comparing the intervention with anything.

\begin{table}[htbp]
  \centering
  \caption{PICOC terms}
  \label{tab:picoc}
  \begin{tabular}{ccp{9cm}}
    \hline
    PICOC & Keywords & Synonyms \\
    \hline
    Population & chatbot &	
    chatterbot,  
    conversational agent,  
    conversational interface,  
    conversational system,  
    dialogue system,   \\
    Intervention  &interaction 	&
    conversation,  
    expectation,  
    experience,  
    impact,  
    perception,  
    usability,  
    user journey \\
    Comparison & \textit{Not applicable} & \textit{Not applicable} \\
    Outcome & satisfaction 	&
    accept,  
    content,  
    effective,  
    enjoy,  
    happiness,  
    preference,  
    quality,  
    trust \\
    Context & text-based & 
        not embodied,   not speech,   not spoken  \\
  \hline
\end{tabular}
\end{table}

We first defined our main keywords for the PICOC: \textit{chatbot}, \textit{interaction}, \textit{satisfaction}, and \textit{text-based}. Then, we did exploratory research with synonyms defined by us in order to see if the results were relevant and what were the missing synonyms we did not think of. We have used VOSViewer\footnote{https://www.vosviewer.com/} to help with the visualization of the missing synonyms. After some iterations of these steps, we came up with the final adjusted generic string:

\vspace{0.3cm}
\colorbox{formalshade}{\parbox{0.9\columnwidth}{\emph{(chatbot OR chatterbot OR ``conversational agent" OR ``conversational interface" OR ``conversational system" OR ``dialogue system") \textbf{AND} (interaction OR conversation OR expectation OR experience OR impact OR perception OR usability OR ``user journey") \textbf{AND} (satisfaction OR accept OR content OR effective OR enjoy OR happiness OR preference OR quality OR trust) \textbf{AND} (NOT embodied AND NOT speech AND NOT spoken)}}}
\vspace{0.3cm}

The digital databases chosen to run the string were \href{https://dl.acm.org/}{ACM Digital Library}, \href{https://ieeexplore.ieee.org/Xplore/guesthome.jsp}{IEEE Xplore}, and  \href{https://www.scopus.com/}{Scopus} and their respective seach string is available on Zenodo\citep{anonymous_zenodo}. They were chosen for being extremely relevant to software engineering research \citep{BreretonKBTK07}, for indexing a great number of conferences and journals, and for being able to run our generic search string directly in its entirety. Table \ref{tab:strings} shows the specific string for each source. 

\begin{table}[htbp]
 \centering
 \caption{Search Strings per Source}
 \label{tab:strings}
 \footnotesize
 \begin{tabular}{p{1cm}p{13cm}}
    \hline
    Source & String \\
    \hline
    ACM Digital Library & [[Abstract: chatbot] OR [Abstract: chatterbot] OR [Abstract: ``conversational agent"] OR [Abstract: ``conversational interface"] OR [Abstract: ``conversational system"] OR [Abstract: ``dialogue system"]] AND [[Abstract: interaction] OR [Abstract: conversation] OR [Abstract: expectation] OR [Abstract: experience] OR [Abstract: impact] OR [Abstract: perception] OR [Abstract: usability] OR [Abstract: ``user journey"]] AND [[Abstract: satisfaction] OR [Abstract: accept] OR [Abstract: content] OR [Abstract: effective] OR [Abstract: enjoy] OR [Abstract: happiness] OR [Abstract: preference] OR [Abstract: quality] OR [Abstract: trust]] AND NOT [Abstract: speech] AND NOT [Abstract: embodied] AND NOT [Abstract: spoken] AND [Publication Date: (01/01/2011 TO *)] \\
    IEEE Xplore & ((``All Metadata":chatbot OR ``All Metadata":chatterbot OR ``All Metadata":``conversational agent" OR ``All Metadata":``conversational interface" OR ``All Metadata":``conversational system" OR ``All Metadata":``dialogue system") AND (``All Metadata":interaction OR ``All Metadata":conversation OR ``All Metadata":expectation OR ``All Metadata":experience OR ``All Metadata":impact OR ``All Metadata":perception OR ``All Metadata":usability OR ``All Metadata": ``user journey") AND (``All Metadata":satisfaction OR ``All Metadata":accept OR ``All Metadata":content OR ``All Metadata":effective OR ``All Metadata":enjoy OR ``All Metadata":happiness OR ``All Metadata":preference OR ``All Metadata":quality OR ``All Metadata":trust) AND NOT ``All Metadata":voice AND NOT ``All Metadata":speech AND NOT ``All Metadata":embodied AND NOT ``All Metadata":spoken)
    \\
    Scopus & TITLE-ABS-KEY ( ( chatbot  OR  chatterbot  OR  ``conversational agent"  OR  ``conversational interface"  OR  ``conversational system"  OR  ``dialogue system" )  AND  ( interaction  OR  conversation  OR  expectation  OR  experience  OR  impact  OR  perception  OR  usability  OR  ``user journey" )  AND  ( satisfaction  OR  accept  OR  content  OR  effective  OR  enjoy  OR  happiness  OR  preference  OR  quality  OR  trust )  AND NOT  voice  AND NOT  speech  AND NOT  embodied  AND NOT  spoken )  AND  PUBYEAR  >  2010  AND  ( LIMIT-TO ( DOCTYPE ,  ``cp" )  OR  LIMIT-TO ( DOCTYPE ,  ``ar" ) )  AND  ( LIMIT-TO ( LANGUAGE ,  ``English" )  OR  LIMIT-TO ( LANGUAGE ,  ``Spanish" )  OR  LIMIT-TO ( LANGUAGE ,  ``Portuguese" ) )  AND  ( LIMIT-TO ( SRCTYPE ,  ``p" )  OR  LIMIT-TO ( SRCTYPE ,  ``j" ) )  \\
  \hline
\end{tabular}
\end{table}

The initial idea was to run the strings for title, abstract and keywords. However, the search engines did not share this specific option. In ACM, it was either the title or the abstract without repeating the string, hence we chose abstract. In IEEE, ``All Metadata" refers to title abstract and keywords as well as ``TITLE-ABS-KEY" in Scopus.
 
We took as an advantage the available filtering and string options for each source to apply some our exclusion criteria and automate the exclusion of unwanted papers as suggested by \citet{CostalFFQ21}. In ACM Library, we applied a manual filter for ``Content Type" selecting only ``Research Article" and selected a date range starting from 2011 until the time of the search. In IEEE Xplore, we applied a manual filter for the date range. In Scopus, all filters were added to the string which were the ``Document Type" as ``Conference Paper" or ``Article"; the ``Publication Year" as older than 2010; ``Language" as ``English" or ``Portuguese" or ``Spanish"; and the ``Source Type" as ``Journal" or ``Conference Proceedings".

\subsection{Selection Criteria}

As suggested by \citet{ barbara2007guidelines}, we have defined both inclusion and exclusion criteria. The inclusion criteria refer to the main theme of the papers and which ones should be accepted in the context of chatbots. The inclusion criteria (IC) are shown below:

\begin{enumerate}
    \item[(IC)] Tests one or more text-based chatbot conversational practices with users by:
    \begin{enumerate}
        \item[(1)] applying practices to real text-based chatbots; 
        \item[(2)] simulating practices through Wizard of Oz experiments or similar;
        \item[(3)] by showing examples of interactions containing the practices and gathering users' impressions;
        \item[(4)] by explaining the idea of the practices to users and gathering their impressions about it.
    \end{enumerate}
\end{enumerate}

The exclusion criteria aim to exclude papers that fit the inclusion criteria but do not present some methodology, focus, aspect, or approach \citep{CostalFFQ21}. The exclusion criteria (EC) are shown below:

\begin{enumerate}
    \item[(EC1)] It does not present the impact of the conversational strategy or the individual practices;
    \item[(EC2)] The object of analysis does not refer entirely to text-based one-on-one interactions between chatbots and humans (e.g. spoken interaction, embodied agent, machine-machine interaction, social media bots, etc);
    \item[(EC3)] The focus is not on conversational design and user interaction;
    \item[(EC4)] It is written in a language other than the ones understood by the authors (Portuguese, Spanish, and English);
    \item[(EC5)] It is not a primary full research paper (e.g. book chapters, magazine articles, dissertation, thesis, literature reviews, work in progress, position paper, duplicated work, etc);
    \item[(EC6)] Published before 2011.
\end{enumerate}

Regarding EC1, it guarantees that the select works properly identified and actively tested or observed the impact of these practices with users, being them either a human sentiment or a behavior pattern. The identified impact can be a result of using a single practice or a set of practices to comply with a wider conversational strategy.

It is important to set apart embodied agents from text-based agents with avatars, as verified by EC2. Embodied agents communicate intentions or messages via body expressions while text-based agents only communicate via text although they can be represented by an image, either human, robotic, or zoomorphic. Simpler and directly, we accepted agents with an icon or figure as long as it is static. Moreover, we are not interested in chatbots that are only ``bots", which we considered as agents that communicate but they are not capable of engaging in a whole conversation with a human. 

If the work is focused on the technicalities of the conversational practice and not user experience, it will be discarded by EC4. This can be easily detected by identifying the variables being measured. For example, when considering sentiment analysis and adaptive responses as conversational practice, if the user study is interested in the accuracy or performance of the algorithm, it will be discarded. On the other hand, if the variables measured in the user study are user satisfaction or other user-centered variables, it will be considered.

The other ECs are related to quality standards to ensure that the selected papers are credible, have a fully developed research, and were properly reviewed. Moreover, it is important to define a short time range to accepted papers since chatbots are emergent and constantly evolving technology, which makes some older works lose relevance, especially in human-AI interaction. Therefore, we discarded studies before 2011 since it was in the early 2010s that chatbots started to get known by the general public \citep{rapp2021}.

\subsection{Quality Assessment}

Even though the exclusion criteria already give us relevant works, it is still necessary to run a quality checklist to ensure that the practices are valid and their impacts were properly and scientifically measured to be considered in our review. For that, we have used the following checklist to accept papers:

\begin{enumerate}
    \item[(QA1)] Are conversational practices pragmatic and replicable? 
    \item[(QA2)] Is the methodology clear, adequate, and well-defined?
    \item[(QA3)] Is the number of participants sufficient for wider inference?
    \item[(QA4)] Is there a comparison or control group testing the violation or absence of the conversational practices?  
    \item[(QA5)] Are results clear and relevant?
    \item[(QA6)] Are the impacts statistically calculated? If not, the qualitative analysis is adequate?
    \item[(QA7)] Are the impacts of conversational practices properly presented and classified as positive, negative, or neutral?
    \item[(QA8)] Are limitations and threats to validity presented?
\end{enumerate}

If a study did not attend all questions of this checklist, it was discarded. Since this work is concerned not only with practices but their impacts on users, this checklist established a rigorous cutting line to guarantee scientifically well-measured results, and this is specially checked by QA2 to QA8. Moreover, QA1 aided the removal of works that approached subjective practices, for example, a work that proposes ``proactive" chatbots but does not test proper applications of it such as the chatbot ``starting the conversation", which is what we are interested in.

\subsection{Conducting}

\begin{figure}[htbp]
  \centering
  \includegraphics[width=0.7\columnwidth]{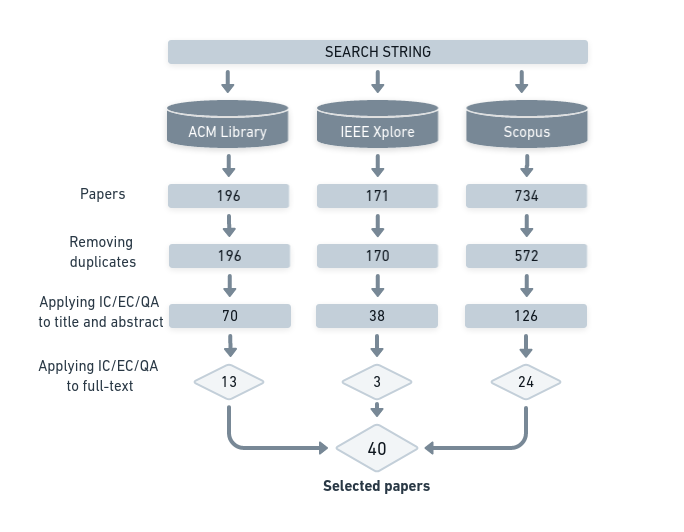}
  \newline
    \footnotesize{IC=inclusion criteria EC=exclusion criteria QA=quality assessment}
  \caption{Remaining papers after each step of the SLR.}
  \label{fig:slr}
\end{figure}

For conducting the review we have used Parsifal\footnote{https://parsif.al}, which is a free open source web platform for supporting SLR. Parsifal was chosen because its features and workflow were based on the SLR process used in this work that was proposed by Kitchenham and Charters \citep{barbara2007guidelines}. It streamlines review by providing easy and fast navigation through titles and abstracts during filtering and automatic detection of duplicated papers. Figure \ref{fig:slr} shows the remaining papers after each step of the review.

We collected studies through the search string until February 2022. We had to remove a lot of duplicates since Scopus indexes work from ACM Library and IEEE Xplore as well. Then, we applied the inclusion and exclusion criteria by reading the title and abstract. In general, all works were about chatbots, but many did not approach conversational practices or did not have user studies. Lastly, we read the full text for deeper analysis and application of quality assessment. Figure \ref{fig:slr} shows the remaining papers after each step of the review and the detailed dataset is available on Zenodo at \href{https://doi.org/10.5281/zenodo.6399190}{https://doi.org/10.5281/zenodo.6399190} \citep{anonymous_zenodo}.

\subsection{Data Extraction}\label{sec:data_extraction}

In systematic reviews, data extraction is vital for building quantitative and straightforward views of the set of studies. However, our review has the objective of serving as a basis for the construction of a guide, therefore being quite restrictive as seen in IC/EC/QA and it does not tend to have a broad view of the area. Accordingly, the categories of information to be extracted, which are shown in Appendix \ref{ap:data_extracted}, are highly tied to the research questions. Apart from this reasoning, Table \ref{tab:selected} presents the year of each paper for tracking the evolution of works and the number of users that participated in the paper's experiment, since the higher the sample the higher the relevance of impacts.

Regarding RQ.1, we have extracted data for the column ``Conversational practice(s)" as excerpts from selected papers, therefore, there is the use of several terms for the same practice (e.g. dynamically delayed responses and adaptive response speed) to preserve the integrity of the extracted information. Additionally, there were many papers that used more than one practice to achieve a goal, thus we extracted these goals to fill the column ``Strategy". For this column, we did open coding to assign a strategy to each paper, which is the process of iteratively working on a set of concepts that will later be grouped and classified \citep{Wolfswinkel2013}.

The column ``Impact(s)" was extracted to support the discussion of RQ.2. Although this column was also filled with excerpts, it was adapted for better understanding. If the identified impact referred to a user's feeling or behavior, we filled it with the excerpt itself (e.g. enjoyment, satisfaction, and self-disclosure). On the other hand, if the impact that the paper evaluates refers to how the user perceives a chatbot attribute, we added the expression ``perception of" before it. The impacts refer to the combined use of the listed practices, except in the cases explained in the footer of Table \ref{tab:practices}.

Lastly, columns ``Moderator(s)" and ``Context" support the discussion of RQ.3. The moderators were extracted exactly as originally written in the paper and refer to variables that reduce, empower or change how the impacts of practices. These variables were either measured statistically with a significant result or they were observed by authors in their experiments and presented with qualitative analysis. The chatbot's context is also an indirect moderator since after the data extraction it is possible to check for different impacts of the same practices in different domains. The context was also extracted through open coding.

\subsection{SLR Results}

This section describes the process and results of the SLR's data extraction and answers the research questions by interpreting these results. It also discusses the implications of implementing the conversational practices considering the aggregated result of the selected studies, considering that some of them address the same practices. 

\subsubsection{RQ.1. What are the textual or visual approaches used in chatbot conversational design?}

There are a lot of practices used in the literature with the intention of causing good impressions on users, although some practices ended up having a negative or neutral impact in some works. These practices are mostly used with the aim of humanizing the chatbot, and they make use of different visual, linguist, or interactive elements. 

Regarding visual elements, avatars stand-out as a straightforward way of making the chatbot look literally human \citepPS{DeCicco20201213, Liao2020430, Toader2020}, however, defining a chatbot avatar is not as easy as it seems because its gender and looks can cause impressions in users even before they interact with the chatbot \citepPS{Toader2020}. Other visual elements are emojis or emoticons \citepPS{Beattie2020409, DeCicco2021140, Fadhil2018}, which makes an interplay with linguistic elements since they accompany or substitute text messages to convey feelings and emotions. As well as emojis, GIFs, and memes \citepPS{DeCicco2021140, Tsai2021460} can be used to enhance the expressiveness of emotions.

On the other hand, there is a wide variety of linguist elements used to enhance chatbot conversations, ranging from message content to message formatting, which can convey personality traits coming from the chatbot. For example, small talk is a practice of talking about casual and out-of-domain subjects \citepPS{Lee2020, DeCicco2021140}, such as greetings, jokes, and the chatbot's fictional background. For that, self-disclosing the chatbot's non-human identity plays a vital role since the decision of revealing the chatbot's true identity as a virtual agent \citepPS{Mozafari20212916, Mozafari20214} will define its background. 

Moreover, chatbots can be emphatic by adapting responses to what has been said by the user \citepPS{Diederich2019} such as demonstrating sadness when the user informing something went wrong or by echoing users responses through reaffirming what the user has just said \citepPS{Rhim2022}. These practices can also be applied when the chatbot cannot solve a problem or does not understand a message as a means of repairing a breakdown \citepPS{Ashktorab2019}. Another way of conveying feeling through messages is by leveraging on punctuation, such as exclamatory feedback \citepPS{DeCicco2021140}. Moreover, jokes and funny stories can also be used to pass on joy and excitement \citepPS{Ceha2021}.

The initial message of the chatbot can be decisive for user retention. Therefore, chatbots can start by presenting themselves as a human or a machine \citepPS{Rhim2022, Mozafari20212916, Zhou2019}, telling its name \citepPS{Liao2020430, Schanke2021736}, welcoming the user with a friendly message \citepPS{Kull2021840} and presenting what are its capabilities \citepPS{Mozafari20214}. This moment is also adequate for collecting the user name for later use when addressing the user during conversation \citepPS{Rhim2022, Tsai2021460}. In the specific case of dealing with bilingual users, the chatbot can also code-mix, which consists of inserting foreign words in the middle of the message \citepPS{Bawa2020}.

In terms of language choices, chatbots can be polite \citepPS{RanaMS21} but with some touches of well dosed informality \citepPS{Svikhnushina2021}. Language can also be personalized to match the domain by varying the use of verbs, pronouns, conjunctions and other linguistic features \citepPS{Chaves2022}.  Moreover, chatbots can mirror their users' language style \citepPS{Spillner2021}.

Chatbots can help users to understand them more by being honest and open with what happens behind the conversation. Acknowledging misunderstanding and suggesting solutions are ways of leading users out of a conversation breakdown \citepPS{Ashktorab2019}. Moreover, justifications and explanations are specially important for users to understand why they are receiving some information, instruction or recommendation from the chatbot \citepPS{Pecune2022, Wilkinson2021}.

Lastly, interaction elements can be used to humanize the agent, streamline conversation and avoid breakdowns. Typing cues and dynamic delayed responses increases the impression that there is a human being typing at the other side by masking the instantaneous response of the chatbot \citepPS{Schanke2021736, Gnewuch2018}. Beyond that, buttons and carousels are useful for guiding users to quickly send the message the chatbot expects to well function \citepPS{Diederich2020}.

\subsubsection{RQ.2. What are the positive or negative impacts on users of the identified textual or visual approaches in chatbot conversational design?}\label{sec:rq2}

The impacts investigated by the selected papers are tied up with the chatbot domain. Thus, works that investigate the impacts of chatbots in the health context are more interested in sentiments that empower users' well-being such as motivation and enjoyment \citepPS{Fadhil2018, Liao2020430}. For mental health, it is important that the user develops feelings of closeness, friendship, and trust in order to self-disclosure to the chatbot \citepPS{Lee2020}, which is an important factor for the success of mental treatment. Conversely, commercial brands are more interested in user retention and satisfaction ratings, although the other feelings previously cited are not an impediment to the success of customer service. 

It is no secret that users prefer a human agent rather than a virtual agent \citep{Lei20213977}. Because of that, a chatbot that discloses itself as non-human can cause users to instantly lose trust \citepPS{Mozafari20212916}, which is not surprising since it is natural that humans have higher confidence towards other humans rather than machines. However, this effect is not due to the displayed identity but to the perceived identity \citepPS{Shi2020}, that is, the identity that users believe is the true one. 

If the chatbot pretends it is human and the user is suspicious about it, the negative impact may be a lot worse than disclosing the chatbot identity as non-human because the user will feel deceived \citepPS{Schanke2021736}, get angrier, and more frustrated \citep{GRIMES2021113515}. Moreover, disclosing identity combined with showing what the chatbot is capable of as well as communicating it's weaknesses can produce trust levels corresponding to that of undisclosed conversational partners \citepPS{Mozafari20212916}. Therefore, for this specific practice of disclosing identity, we can consider that the decrease in perception of humanness after disclosure \citepPS{Hendriks2020271} is a natural effect that must be endured to avoid worse impacts of users finding out they are being deceived.  

Moving forward, elements used towards humanization, social presence or for conveying emotions have positive impacts on users. These elements enhance users' perception of social presence and anthropomorphism \citepPS{DeCicco2021140, Rhim2022}. For customer service and recommendation chatbots, clarity and openness are more approached by means of an honest self-introduction \citepPS{Mozafari20214} and justifying the chatbot's behaviour \citepPS{Pecune2022, Wilkinson2021}, which are practices well received by users.

Concerning avatars, the impacts found by works are conflicting, ranging from negative, neutral, and positive. \citetPS{Tsai2021460} found that a human name in conjunction with a human avatar only boosts other humanization elements but are not sufficient to impact consumer response. In \citetPS{Pizzi2021878}, the enforcement of anthropomorphism through avatars in automatically activated agents negatively impacted users, whereas the lack of the avatar was positive. These works show that the impact of avatars is highly dependent on other variables.

Still, if avatars are used, gender and physical attributes can improve user perception. Woman avatars make users more satisfied and likely to forgive chatbot errors \citepPS{Toader2020}. Racial identity through physical attributes and name of the avatar has a positive impact on closeness, self-disclosure, and satisfaction \citepPS{Liao2020430}, which can be beneficial in both health and commercial contexts. 

Elements of informality, such as casual language, emojis, GIFs, and jokes are coupled with sentiments of joy, closeness, friendship, motivation in health and learning contexts but are not determining user satisfaction in commercial contexts \citepPS{DeCicco2021140, Ceha2021, Fadhil2018}. However, since politeness is also a positive element \citepPS{RanaMS21, Svikhnushina2021}, the use of informality must be moderated so as not to exceed the limits of good manners and to cancel the professional image of the chatbot.

Typing errors were also investigated as an element of informality in an attempt to make the chatbot more human but it did not have a positive impact since users thought it was ``a lack of developer competence" \citepPS{Buhrke20214456}. Interaction elements, such as buttons, have a neutral effect on satisfaction but decrease the perception of humanness, although they can streamline conversations. Therefore, such elements must be used with caution.

\subsubsection{RQ.3. Are there moderating effects of other variables on the identified impacts of practices? }

We have identified different moderating variables that were statistically measured and significantly changed the impacts of the practices. Regarding the interplay among practices, avatars brought a negative impact when using automatic activation \citepPS{Pizzi2021878}; self-defeating jokes (vs affiliative jokes) reduced the positive impacts of humor \citepPS{Ceha2021}; acknowledging limitations smoothed the negative impact of revealing the chatbot identity \citepPS{Mozafari20212916}; revealing identity right away in critical services is detrimental, but right after a failure is positive \citepPS{Mozafari20214}; and pretending to be human decreased positive effects of humanization elements \citepPS{Schanke2021736}.

One practice that particularly suffers from the interplay with other measures is avatars. As we started discussing in RQ.2, human avatars potentiate other humanization techniques but increase the expectation of users. Therefore, users can feel deceived by the human image and be more frustrated with chatbot failures. Moreover, every detail of the human avatar, such as gender and physical attributes, moderate the impact on users. By a joint analysis of selected works, the safest approach seems to be a woman human avatar, with racial mirroring in conjunction with self-disclosing the chatbot as non-human for transparency. Furthermore, although the perception of humanness may decrease, a robot avatar can be used as well, without a negative impact on satisfaction \citepPS{Liao2020430}. 

Other moderating variables that are beyond the control of designers must be taken into account when using the practices such as conversation duration \citepPS{Lee2020}, context of use \citepPS{Zhou2019}, user's personality \citepPS{Zhou2019, RanaMS21, AhmadSR21}, language proficiency \citepPS{Bawa2020}, age \citepPS{Wilkinson2021, RanaMS21}, gender \citepPS{Diederich2019, RanaMS21} and experience with technology \citepPS{Wilkinson2021, Ashktorab2019}. Besides these works and apart from the ones that did not find significant moderating effects from age and gender \citepPS{SHEEHAN202014, Diederich2019, Diederich2020, Gnewuch2018}, others have not tested moderating effects of age and gender. Moreover, the demographics of participants are generally excluding older adults and the elderly as well as adolescents.

In \citetPS{RanaMS21}, it was statistically found that women were more sensitive to the chatbot's polite triggers, being more positively impacted than men. In some cases, men even gave a lower rating for the polite chatbot. In consonance, \citepPS{Chaves2022} found that the correct use of linguistic features (e.g. verbs, coordinate conjunctions, pronouns) positively impacts users, but the linguistic features to be used change from domain to domain. Considering that it is not always possible to run deep tests with target users, one strategy to mitigate these effects is to balance the use of practices to avoid exaggerations. For example, if a chatbot is extremely informal and makes use of jokes, women may feel uncomfortable whereas men feel joyful.

Ahmad \citetPS{AhmadSR21} found that is no correlation between specific user's personality traits and chatbot preferences because the conjunction of many personality traits makes users unique, and they conclude there is a need for personalization during the conversation. Personalization is limited in conversation design but can be enhanced on the technological side by identifying users' traits with natural language processing, which is not in our scope. However, the previous recommendation of no exaggerating applies, which prevents the chatbot from being inappropriate for certain groups of people.

Designers should also check for users' perceptions of practices. As seen in \citetPS{Ng2020190} and \citetPS{Shi2020}, regardless of the practice that has been used, what matters is what the user believes and perceives, which moderates impacts. If the chatbot is designed to pretend it is human but the user does not believe in this identity, the impact of humanness is annulled. This can be mitigated by running pre-tests to measure the perception of users over practices.

Through qualitative analysis of participants' responses, \citetPS{Ashktorab2019} and \citetPS{Rhim2022} have found that the positive effect of textual practices are harmed by the lack of variability in messages causing users to see the chatbot as ``an auto-machine". Therefore, when using any practice, designers should conceive different responses for conveying the same message.

\section{Proposed Guidelines for Chatbot Conversational Design}

Studying the papers selected by SLR enabled us to propose guidelines to conversational practices that can help designers in building user-centered chatbots. The first step was to build a conceptual map that synthesizes the joint knowledge of the selected studies and it is further explored in Section \ref{sec:conceptual_model}. Based on the conceptual map, we built a simple and objective guide in web page format that explains conversational practices and in which situations they should be used, which is further explained in Section \ref{sec:guide_structure}. 

\subsection{Conceptual Map}\label{sec:conceptual_model}

In the SLR, we kept the extracted data as close as possible to its origin in the paper, as explained in Section \ref{sec:data_extraction}. Answering research questions by looking at the data in Appendix \ref{ap:data_extracted} paved the way to a deeper analysis, which was done through open coding, axial coding, and selective coding. Open coding was already applied and explained in Section \ref{sec:data_extraction}, whereas the other two are responsible for grouping codes into categories and finding interrelationships among them, respectively \citep{Wolfswinkel2013}.

\begin{figure*}[!bhtp]
  \centering
  \includegraphics[width=\textwidth]{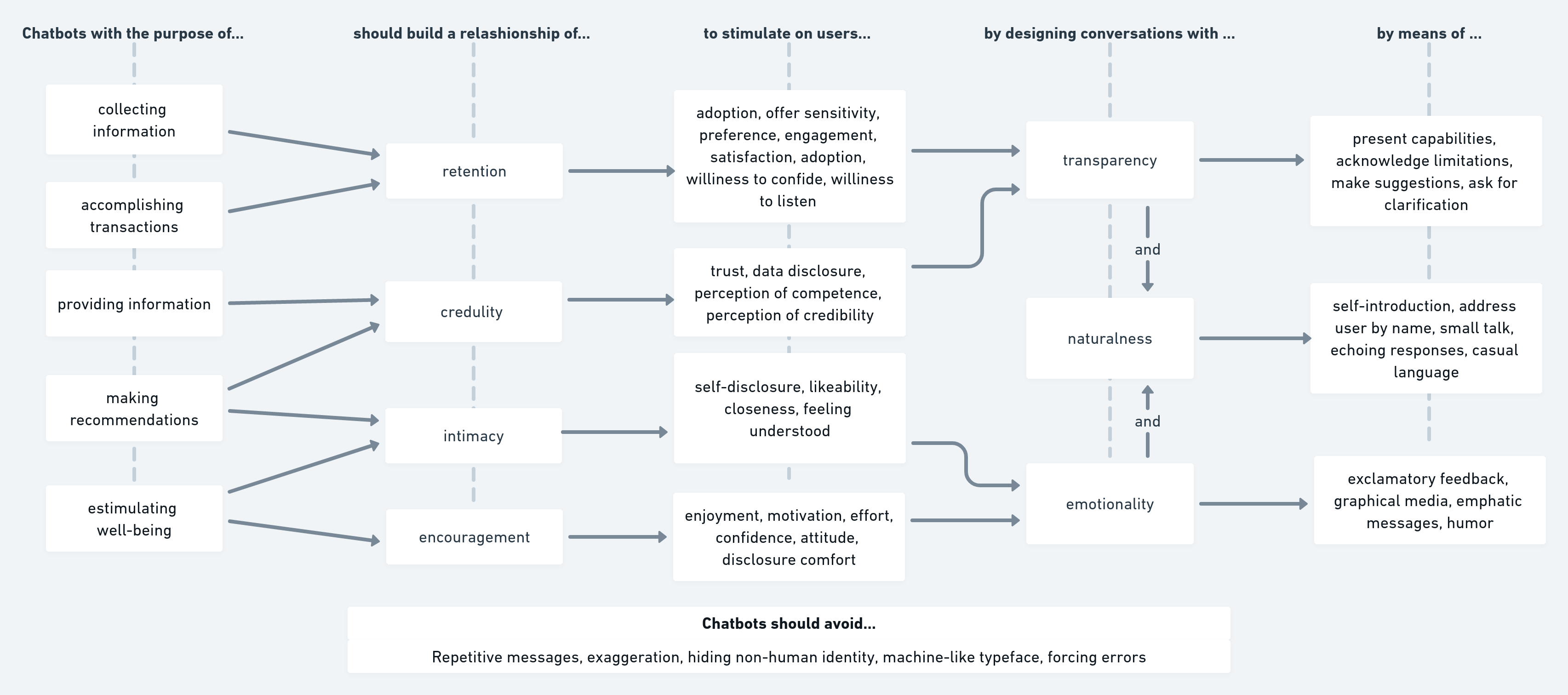}
  \caption{Conceptual map of chatbot conversation design according to purpose and user impact.}
   \label{fig:model}
\end{figure*}

This analysis revealed some patterns of study focus depending on the purpose of the chatbot. For example, papers that tested customer service chatbots were highly interested in satisfaction, whereas papers that tested chatbots in the context of health were more interested in feelings of well-being. These patterns are listed in the conceptual map that is shown in Figure \ref{fig:model}. 

This model starts by separating the purpose of chatbots and linking them to what should be the type of relationship that has to be built with the user. The listed relationships arose from a qualitative and joint analysis of selected papers and are linked to a group of impacts that were investigated by selected papers. From our analysis, it was possible to identify the focus of each group of impacts, which were \textit{transparency}, \textit{naturalness}, and \textit{emotionality}. 

Chatbots that seek to collect information, such as chatbots that conduct surveys and interviews, must keep the attention of users, make them complete the questionnaire, and provide reliable information. Similarly, chatbots that accomplish transactions, such as booking and shopping, must keep their users interested in order to make them complete the purchase. Therefore, to \textit{retain} users, the best approach is to acknowledge capabilities and limitations and set user expectations right away.

\textit{Credulity} is an important factor for users that are looking for information and recommendation in a chatbot since the reliability of responses impacts the trust of users. Moreover, to make recommendations to users, chatbots must know more about them, and this can be achieved by creating a relationship of \textit{intimacy}. Lastly, for stimulating well-being, a chatbot must act as a companion and build a relationship of \textit{intimacy} and \textit{encouragement}.

We have identified three strategies for conversational design that can be used to build these relationships. The set of selected studies revealed that \textit{naturalness} is essential in all of these use cases. \textit{Transparency} is more important when the chatbot needs to be competent and effective, whereas \textit{emotionality} is more important when a deeper connection is necessary to accomplish the chatbot's purpose. 

\textit{Emotionality} and  \textit{transparency} are not mutually exclusive, but designers should switch focus on achieving the desired relationship. For example, users talking to a therapist chatbot are not concerned about being aware of everything of the chatbot, but they are interested in being listened to, understood, and cheered up. In this use case, focusing on being transparent more than being emphatic is detrimental to the user experience, therefore, \textit{Emotionality}'s practices should be enforced and practices \textit{Transparency}'s practices, if used, should not be in the spotlight.

Regarding what \textit{Chatbots should avoid}, \textit{machine-like typeface} came from the findings of \citetPS{Candello2017}, and forcing errors from findings of \citetPS{Buhrke20214456}. \textit{Repetitive messages} came from deeper discussions of participants perceptions in the works of  \citetPS{Ashktorab2019} and \citetPS{Rhim2022}. \textit{Exaggeration} should be avoided because users' personality and demographics have been recurring moderators in selected works, therefore, it is important to stay in the middle ground to please the greatest number of users, especially in the case where designers cannot afford to do extensive user studies before designing the conversations. Lastly, although disclosing the chatbot's true identity results in negative impacts on users' trust, as discussed in Section \ref{sec:rq2}, \textit{hiding non-human identity} has worse effects than disclosing it when the user finds out it is being deceived.

Although the open coding was responsible for reducing the list of practices seen in Table \ref{tab:practices} to the one shown in the conceptual map, some of them do not appear in it intentionally because of conflicting results among papers or little positive impact, such as buttons and avatars. Moreover, some practices such as code-mix, racial mirroring and linguistic alignment are very tied to specific contexts or depend on their audience characteristics, therefore they were not included since the conceptual map intends to be context-independent.

\subsection{Guide Structure}\label{sec:guide_structure}

The proposed guide was developed as a web page that brings in a more accessible language the concepts that were defined in the conceptual map. It is called \textit{Guidelines for Chatbot Conversational Design} (GCCD) and it is entirely avaiable on Zenodo at  \href{https://doi.org/10.5281/zenodo.6399190}{https://doi.org/10.5281/zenodo.6399190} \citep{anonymous_zenodo} alongside other supplementary material. It is composed of the following pages: \textit{Home}, \textit{Conversational Design}, \textit{Naturalness}, \textit{Emotionality}, \textit{Transparency} and \textit{What to avoid}.

The \textit{Home} page presents the justification of the guide and a summary of the guide's contents . The ``Conversational Design" page follows the same structure of the home page, with a text approaching the importance of a well-executed design and the explanation of a shorter version of the conceptual map, with the objective of explaining to users that the practices in the next pages are more relevant in certain contexts. 

\begin{figure*}[!bhtp]
  \centering
  \includegraphics[width=\textwidth]{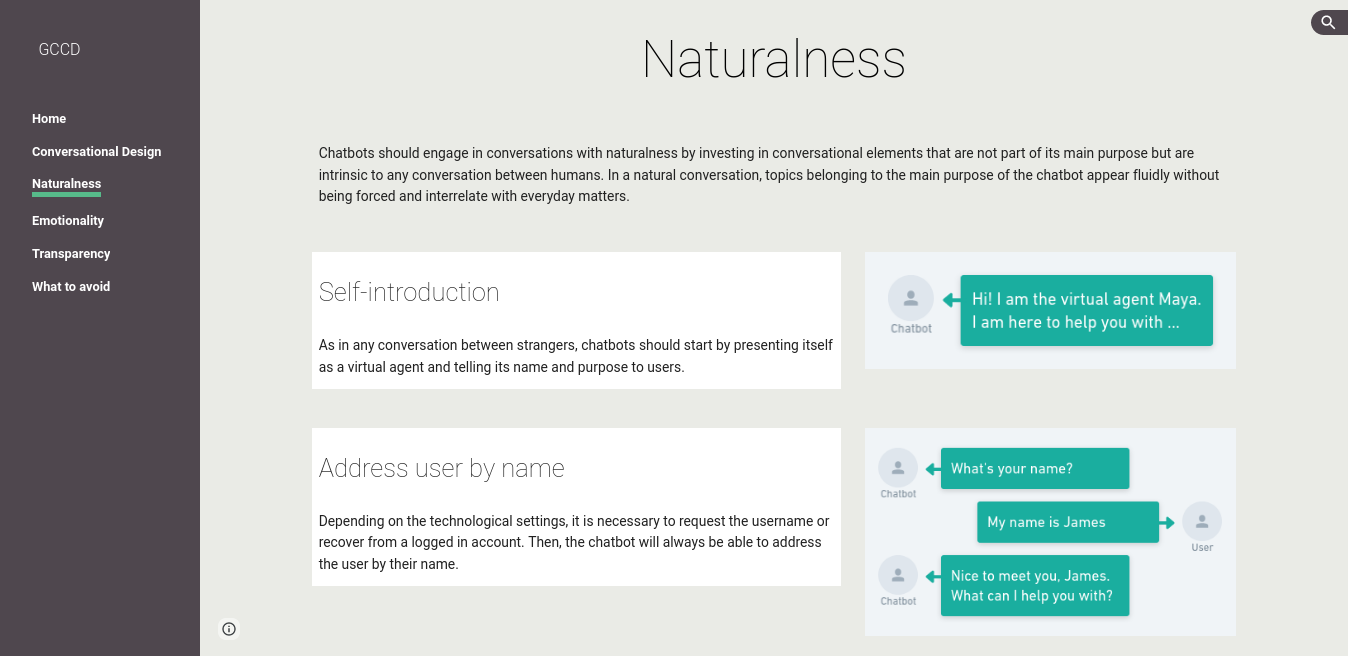}
  \caption{User's view of a guide's page that exemplifies practices for a specific focus, in this case, naturalness.}
   \label{fig:naturalness}
\end{figure*}

On the other hand, the \textit{Naturalness}, \textit{Emotionality}, \textit{Transparency}, and \textit{What to avoid} have a different structure, as shown in Figure \ref{fig:naturalness}. They start with a short paragraph approaching in broad outlines how to achieve the characteristic that entitles the page. Then, for each one of the listed practices, there is a short explanation followed by a figure showing a generic example of a conversational practice that should be implemented or avoided.

All pages have a direct and simple language with the objective of being a practical reference for practitioners with none to advanced knowledge of chatbot development. The guide aims to be an accessible reference for designing effective chatbot conversations even though it is based on a joint analysis of scientific studies with strong theoretical foundations. It is important to note that it only approaches conversational design guidelines, not technological implementations or design processes, as it is out of our scope.

\section{Validation of the Proposed Guidelines}\label{sec:validation}

This section details the methodology used to validate the proposed guidelines, which was carried out through a survey to validate the guide's quality according to technology practitioners, as well as the discussion of the participants' responses and improvements for a future version of the guidelines.  

\subsection{Survey Settings}

The questionnaire was built and distributed through the Google Forms platform, as seen in  \href{https://doi.org/10.5281/zenodo.6399190}{https://doi.org/10.5281/zenodo.6399190} \citet{anonymous_zenodo}, and required a time of 15 to 20 minutes to complete, considering the reading of the guide as a requirement for its completion. Participants were recruited primarily through personal contacts who were technology practitioners. Then, we shared the invitation on social networks and email lists, emphasizing that the survey was aimed at practitioners in the technology area.

The target audience was also reinforced in the terms of consent, which also established that participation was anonymous, voluntary, and with the exclusive purpose of contributing to the success of the research, in addition to the fact that the responses collected could be stored in perpetuity, which could be used anytime for journal publications, conferences, and blog posts. Moreover, they could leave the survey at any time before clicking the send button without any discomfiture since the process of responding was unsupervised. An email was provided in case participants had any problems or questions to the researchers.

The survey was conceived initially in Portuguese for being the mother language of researchers and, consequently, their personal contacts. However, it was also shared in English to reach a wider public. It was only necessary for the participant to select the language on the first page of the survey so that the following pages would appear in the selected language. Besides agreeing to the terms of consent, participants had to confirm that they have read the guide completely before proceeding to questions about it, therefore, we also developed versions of the guide in Portuguese and English, and the proper link would appear to participants according to the language they have chosen. Both versions are available on Zenodo at  \href{https://doi.org/10.5281/zenodo.6399190}{https://doi.org/10.5281/zenodo.6399190} \citep{anonymous_zenodo}.

\subsection{Survey Questions}

The survey had questions approaching practitioners' general experience, experience with chatbots, and their perception of the guide's usefulness and ease of understanding. The original questionnaire from Google Forms is available on Zenodo at  \href{https://doi.org/10.5281/zenodo.6399190}{https://doi.org/10.5281/zenodo.6399190} \citep{anonymous_zenodo}. For practicality, the wording of the survey questions are transcribed below:

\begin{enumerate}
    \item[(Q1)] What is your educational level?
    \item[(Q2)] What is your current main occupation?
    \item[(Q3)] Have you ever researched or worked with chatbots?
    \item[(Q4)] Are you currently researching or working with chatbots?
    \item[(Q5)] What is your level of experience or knowledge of chatbot development?
    \item[(Q6)] Mark how much you agree with each statement [usefulness].
    \begin{itemize}
        \item[(QU1)] Using GCCD would enable me to design a chatbot more quickly.
        \item[(QU2)] Using GCCD would make it easier to design chatbots.
        \item[(QU3)] Using GCCD would make me design chatbots that induce greater user satisfaction.
        \item[(QU4)] Using GCCD would make me design chatbots that induce greater user engagement.
        \item[(QU5)] I would use GCCD for designing a chatbot.
    \end{itemize}
    \item[(Q7)] Mark how much you agree with each statement [ease of use].
    \begin{itemize}
        \item[(QEU1)] I find GCCD easy to use.
        \item[(QEU2)] I find GCCD clear and understandable.
        \item[(QEU3)] I find GCCD flexible to be used with chatbots from different domains.
        \item[(QEU4)] I consider that GCCD requires a lot of knowledge about chatbots to be understandable.
    \end{itemize}
    \item[(Q8)] In your opinion, what are the strengths of GCCD?
    \item[(Q9)] In your opinion, what are the weaknesses of GCCD?
    \item[(Q10)] Is there anything you would change in GCCD? If yes, please explain.
\end{enumerate}

As seen in \href{https://doi.org/10.5281/zenodo.6399190}{https://doi.org/10.5281/zenodo.6399190} \citet{anonymous_zenodo}, from Q1 to Q5 respondents were questioned about their profile and they had to select only one from a number of pre-defined options. These questions were included to verify the diversity of the sample regarding practitioners' general experience and the roles they assumed or could assume in chatbot development. From Q8 to Q10, respondents had an open field at their disposal for full answers, which can help to understand deviant values in closed questions, if necessary.
 
Q6 and Q7 were composed of a set of statements in which users should opt for one number ranging from 1 to 5 that represented a Likert scale of agreement (i.e. strongly disagree, disagree, neither agree nor disagree, agree, and strongly agree). These statements were based on the Technology Acceptance Model (TAM) proposed by \citet{davis1989perceived}, which is a model that measures the degree to which a person believes that using the guide will improve their performance (usefulness) and that it will not involve effort (ease of use). 

\subsection{Results}

The survey collected a total of 66 responses of which 4 came from the English version and the rest from the Portuguese version. All answers are available on Zenodo at  \href{https://doi.org/10.5281/zenodo.6399190}{https://doi.org/10.5281/zenodo.6399190} \citep{anonymous_zenodo}. Figure \ref{fig:profile} depicts the profile of participants according to their responses to questions Q1 to Q5, in which Q4 and Q5 were condensed in the chart (d). It is possible to notice that the sample is very diverse regarding respondents' educational level and main occupation at the time of response. However, regarding experience with chatbots, many of them have not experienced chatbot development, and from the ones that have experienced, most of them have acquired only basic knowledge. Still, we have a relevant amount of respondents with intermediate or advanced knowledge representing more experienced practitioners.


\begin{figure}
\centering
\subfloat[educational level]{%
        \begin{tikzpicture}
            \pie[radius=0.9, text=legend, sum=auto]{24/High School, 21/College graduate, 15/Master's degree, 6/Doctoral degree}
        \end{tikzpicture}
}\hspace{5pt}
\subfloat[current main occupation]{%
        \begin{tikzpicture}
            \pie[radius=0.9, text=legend, sum=auto]{2/Designer, 12/Researcher, 16/Developer, 19/Student, 5/Product Owner, 12/Other }
                \pie[radius=0.8, sum=auto, sum=66, hide number]{2/2}
        \end{tikzpicture}
}\hspace{5pt}
\subfloat[working with chatbots]{%
        \begin{tikzpicture}
            \pie[radius=0.9, text=legend, sum=auto]{34/Never, 23/Previously, 9/Currently}
        \end{tikzpicture}
}\hspace{5pt}
\subfloat[knowledge about chatbots]{
   
        \begin{tikzpicture}
            \pie[radius=0.9, text=legend, sum=auto]{2/Advanced, 10/Intermediate, 31/Basic, 23/None }
                \pie[radius=0.8, sum=auto, sum=66, hide number]{2/2}
        \end{tikzpicture}
    }
\caption{Profile of survey respondents.}\label{fig:profile}
\end{figure}
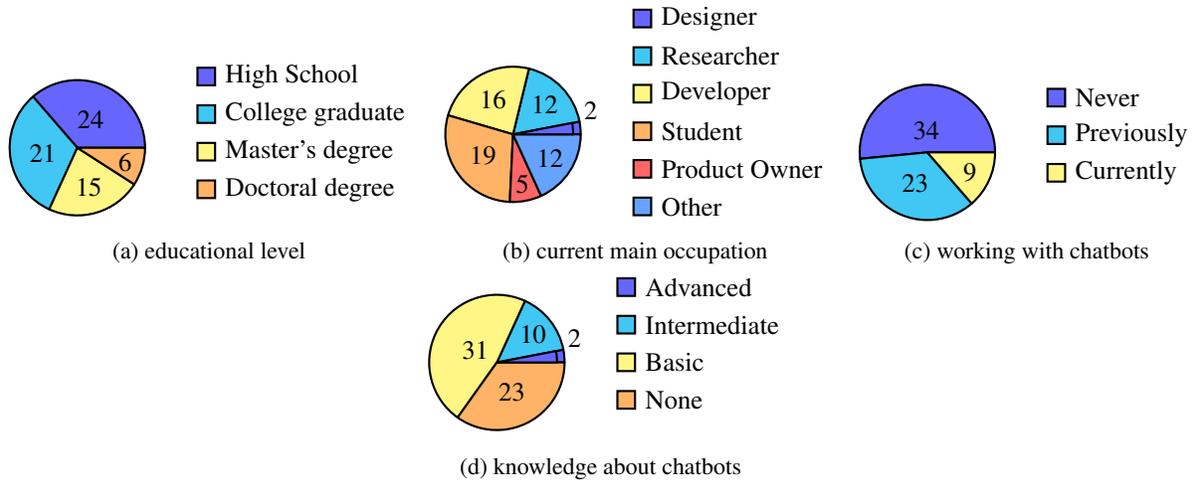
\begin{figure}[!htbp]
    \centering
    \begin{tikzpicture}
        \begin{axis}[
            ybar stacked,
        	bar width=15pt,
        	nodes near coords,
        	visualization depends on={ 
                ifthenelse(meta>2, 0pt, (-1)^\plotnum * 10pt) \as \myshift
            },
            nodes near coords style={ 
                xshift=(\myshift),
            },
            legend style={at={(0.5,-0.20)}, anchor=north,legend columns=-1},
            ylabel={Number of respondents},
            symbolic x coords={QU1,QU2,QU3,QU4,QU5},
            xtick=data,
        ]
        \addplot+[ybar, black, fill=red!80] plot coordinates {(QU1,4) (QU2,3) (QU3,0) (QU4,1) (QU5,1)};
        \addplot+[ybar, black, fill=red!30] plot coordinates {(QU1,9) (QU2,4) (QU3,2) (QU4,2) (QU5,3)};
        \addplot+[ybar, black, fill=yellow!30] plot coordinates {(QU1,16) (QU2,10) (QU3,6) (QU4,8) (QU5,6)};
        \addplot+[ybar, black, fill=blue!30] plot coordinates {(QU1,23) (QU2,31) (QU3,25) (QU4,27) (QU5,21)};
         \addplot+[ybar, black, fill=blue!60] plot coordinates {(QU1,14) (QU2,18) (QU3,34) (QU4,28) (QU5,35)};
        \legend{\strut strongly disagree, \strut disagree, \strut neutral, \strut agree, \strut strongly agree}
        \end{axis}
    \end{tikzpicture}
    \caption{Respondents' level of agreement about GCCD'S usefulness to (QU1) quicken design; (QU2) facilitate design; (QU3) induce greater user satisfaction; (QU4) induce greater user engagement; and (QU5) if they would use it. }
    \label{fig:tam_usefulness}
\end{figure}
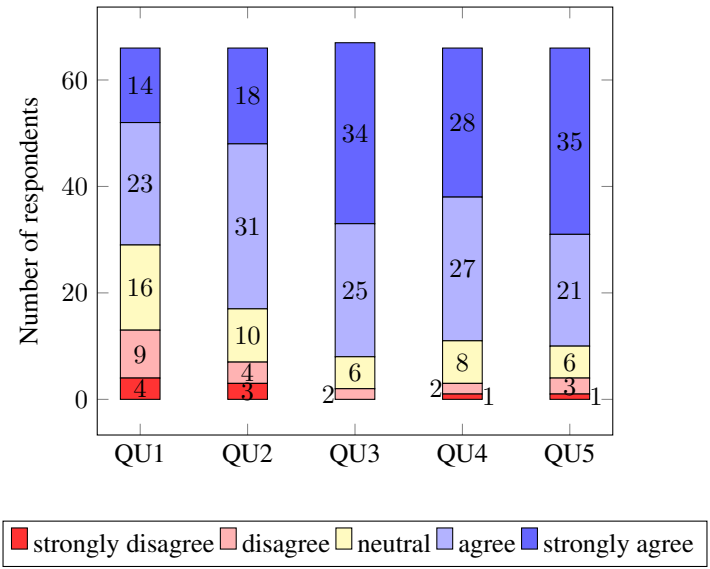
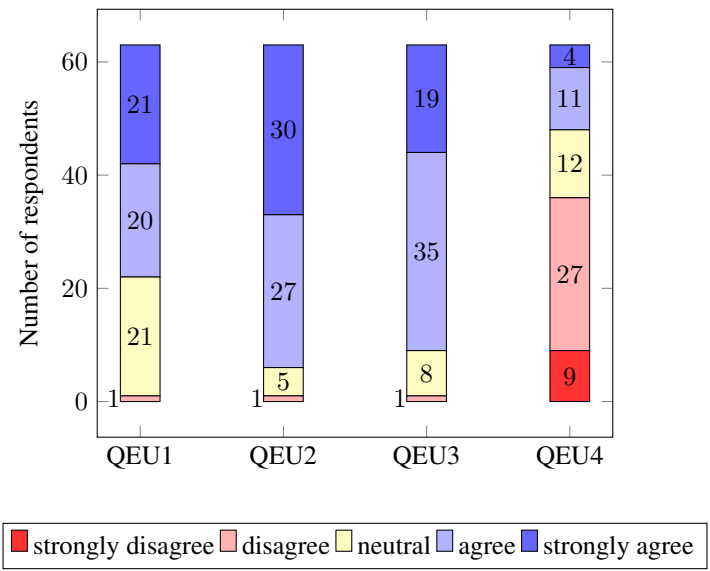
\begin{figure}[!htbp]
    \centering
    \begin{tikzpicture}
        \begin{axis}[
            ybar stacked,
        	bar width=15pt,
        	nodes near coords,
        	visualization depends on={ 
                ifthenelse(meta>2, 0pt, (-1)^\plotnum * 10pt) \as \myshift
            },
            nodes near coords style={ 
                xshift=(\myshift),
            },
            legend style={at={(0.5,-0.20)}, anchor=north,legend columns=-1},
            ylabel={Number of respondents},
            symbolic x coords={QEU1,QEU2,QEU3,QEU4,QEU5},
            xtick=data,
        ]
        \addplot+[ybar, black, fill=red!80] plot coordinates {(QEU1,0) (QEU2,0) (QEU3,0) (QEU4,9)};
        \addplot+[ybar, black, fill=red!30] plot coordinates {(QEU1,1) (QEU2,1) (QEU3,1) (QEU4,27)};
        \addplot+[ybar, black, fill=yellow!30] plot coordinates {(QEU1,21) (QEU2,5) (QEU3,8) (QEU4,12)};
        \addplot+[ybar, black, fill=blue!30] plot coordinates {(QEU1,20) (QEU2,27) (QEU3,35) (QEU4,11)};
        \addplot+[ybar, black, fill=blue!60] plot coordinates {(QEU1,21) (QEU2,30) (QEU3,19) (QEU4,4)};
        \legend{\strut strongly disagree, \strut disagree, \strut neutral, \strut agree, \strut strongly agree}
        \end{axis}
    \end{tikzpicture}
    \caption{Respondents' level of agreement about GCCD'S ease of use regarding it being (QEU1) easy to use; (QEU2) clear and understandable; (QEU3) flexible to be used with chatbots from different domains; and (QEU4) it requiring a lot of knowledge about chatbots to be understandable. }
     \label{fig:tam_ease}
\end{figure}

Concerning respondents' perceptions of the guide's usefulness, Figure \ref{fig:tam_usefulness} shows that around 89\% of respondents agreed on some level that the proposed guide would induce greater user satisfaction (QU3) and  83\% that it would induce greater user engagement (QU4). Their perception is lined up with our findings from the SLR since we selected practices with positive impacts on users. Consequently, around 85\% also agreed on some level that they would use the guide to design a chatbot (QU5).

Although all aspects of usefulness had a majority of agreement, QU1 and QU2 were a bit less positive than the other aspects. We analyzed responses from the open questions Q8 and Q9 to try to understand why, and it is clear that some respondents are concerned with the technical difficulties of implementing these guidelines, which goes against quickening and facilitating design, which is measured by QU1 and QU2, respectively. 

Figure \ref{fig:tam_ease} presents respondents' perceptions about the ease of use of the guidelines. Around 86\% of respondents agreed on some level that the guidelines are clear and understandable (QEU2). Similarly, the vast majority also agreed on some level that the guidelines are flexible to be used with chatbots from different domains (QEU3), which confirms that guidelines are broad enough. 

In line with the percentage of agreement regarding ease of use, around 67\% disagreed on some level that it requires a lot of knowledge about chatbots to put the guide into practice (QEU4), meaning that respondents believe that starters would not have problems understanding the guidelines. This confirms that guidelines are clear enough for all audiences.

Still, in Figure \ref{fig:tam_ease}, it is possible to notice in QEU1 that the majority also agreed that the guidelines are easy to use, however, it has a larger amount of neutral responses. This is also a reflection of respondents' concerns about the technical requirements to implement the guidelines, as discussed before, which implies a higher level of difficulty. 

Responses from Q8 to Q10 are written comments about the guide that can help to uncover possibilities for improvement. Starting with the guide's strengths, the examples for each practice were the high point for most of the respondents, since it helps visualize how it can be implemented. Other strengths mentioned were the objectivity, and clearness of the guide, which reinforce the results shown by the TAM questions. 

On the other hand, simplicity was seen as the main weakness from the point of view of most respondents. They missed a deeper approach to conversational design processes and technological implementation of guidelines. Other weaknesses mentioned by participants were the simplicity of the guide's web design and the theoretical nature of the guide that could lead to technical difficulties in implementing it. 

In line with what was said regarding the guide's weaknesses, suggestions for improvement are mostly related to more in-depth examples, applications of the guide, and references of its implementation to give more credibility. Around 20 respondents did not have any suggestions for improvement because they considered that the guide fulfills its purpose or because they did not feel able to contribute. Furthermore, other comments from Q8 to Q10 were isolated or out of the scope of our work. 

Lastly, we calculated the Relative Strength Index to measure the degree of agreement for each TAM question and we run the Fisher’s Exact Test to verify if there are significant differences among responses from participants with different experiences in chatbot development. These tests were chosen based on the work of \citet{SILVEIRA2022111223}, which also used these calculations over a TAM questionnaire to evaluate usability design guidelines for monitoring interfaces. 

As proposed by Wilder \citep{wilder1978new} and adapted by \citet{SILVEIRA2022111223}, the Relative Strength Index (RSI) is shown in Equation \ref{eqn:rsi}, in which \textit{Ag} refers to the frequency of responses of agreement (i.e. agree and strongly agree) and \textit{Dis} to the frequency of responses of disagreement (i.e. disagree and strongly disagree). After calculating the RSI, the results can be labeled according to an interpretation of values \citep{SILVEIRA2022111223}, as seen in Table \ref{tab:rsi}.

For the Fisher’s Exact Test, we divided the participants into two groups based on their answers to Q3 to check if there is a significant difference between responses from participants who have worked or researched with chatbots compared with the ones who did not. The calculation was done through a web tool \citep{vasavada}. Table \ref{tab:fisher} shows that for all of the TAM questions the p-value is above 0.05 indicating that participants' previous experience with chatbots did not have a significant influence on their responses. 

\begin{equation}
RSI = 100 - \frac{100}{\frac{\textit{Ag}}{\textit{Dis}} + 1}
\label{eqn:rsi}
\end{equation}

\begin{table}[htb!]
  \centering
  \caption{Data for the calculation of the degree of the agreement through the Relative Strength Index (RSI) for each TAM question.}
  \label{tab:rsi}
  \begin{tabular}{ccccccccc}
    \hline
    ID & Ag & Dis & RSI & Interpretation  \\
    \hline
        QU1 & 37 & 13 & 74.0 & Moderate agreement   \\
        QU2 & 49 & 7 & 87.5 & Substantial agreement   \\
        QU3 & 59 & 2 & 96.7 & Very strong agreement  \\
        QU4 & 55 & 3 & 94.8 & Very strong agreement  \\
        QU5 & 56 & 4 & 93.3 & Very strong agreement \\
        \hline
        QE1 & 41 & 1 & 97.6 & Very strong agreement \\
        QE2 & 57 & 1 & 98.3 & Very strong agreement  \\
        QE3 & 54 & 1 & 98.2 & Very strong agreement  \\
        QE4 & 15 & 36 & 29.4 & Moderate disagreement   \\
  \hline
\end{tabular}
  \begin{minipage}{7.5cm}
    \centering
    \footnotesize{
        Ag=frequency of responses of agreement \newline Dis=frequency of responses of disagreement
    }
  \end{minipage}
\end{table}

\begin{table}[htb!]
  \centering
  \caption{Data for the calculation of p-value through Fisher’s Exact Test comparing participants with or without previous research or working experience with chatbots.}
  \label{tab:fisher}
  \begin{tabular}{ccccccccc}
    \hline
    ID & Experience & SD & D & N & A & SA & p-value \\
    \hline
        \multirow{2}{*}{QU1} & Yes & 2 & 5 & 11 & 9 & 5 & \multirow{2}{*}{0.34}  \\
            & No  & 2 & 4 & 5 & 14 & 9 &    \\
        \multirow{2}{*}{QU2}  & Yes & 2 & 2 & 5 & 14 & 9 & \multirow{2}{*}{0.97}  \\
            & No  & 1 & 2 & 5 & 17 & 9 &   \\
        \multirow{2}{*}{QU3}  & Yes & 0 & 2 & 3 & 10 & 17 & \multirow{2}{*}{0.45} \\
            & No  & 0 & 0 & 3 & 15 & 16 &   \\
        \multirow{2}{*}{QU4}  & Yes & 1 & 1 & 5 & 10 & 15 & \multirow{2}{*}{0.47}   \\
            & No  & 0 & 1 & 3 & 17 & 13 &   \\
        \multirow{2}{*}{QU5}  & Yes & 1 & 3 & 2 & 11 & 15 & \multirow{2}{*}{0.26}  \\
            & No  & 0 & 0 & 4 & 10 & 20 &  \\
        \hline
        \multirow{2}{*}{QE1} & Yes & 0 & 1 & 12 & 9 & 10 & \multirow{2}{*}{0.58} \\
            & No & 0 & 0 & 9 & 13 & 12 &   \\
        \multirow{2}{*}{QE2} & Yes & 0 & 1 & 2 & 15 & 14 & \multirow{2}{*}{0.87}  \\
            & No & 0 & 0 & 3 & 14 & 17 &    \\
        \multirow{2}{*}{QE3} & Yes & 0 & 1 & 4 & 17 & 10 & \multirow{2}{*}{0.91} \\
            & No & 0 & 0 & 4 & 20 & 10 &   \\
        \multirow{2}{*}{QE4} & Yes & 4 & 12 & 8 & 6 & 2 & \multirow{2}{*}{0.73}  \\
            & No & 6 & 15 & 4 & 6 & 3 &    \\
  \hline
\end{tabular}
  \begin{minipage}{8cm}
    \centering
    \footnotesize{
        SD=strongly disagree  D=disagree  N=neither agree nor disagree  A=agree  SA=strongly agree
    }
  \end{minipage}
\end{table}

\section{Discussion}

This section presents practical and theoretical contributions of the guide based on the results presented in the previous sections. It also discusses the improvements that could be implemented considering the practitioners' comments collected by the survey.

\subsection{Managerial Implications}

Overall, all of the aspects regarding the usefulness and ease of use had positive results considering responses to the TAM questions. Considering the diversity of the respondent's level of education, we can infer that our proposed guide is useful in both academic and industrial settings. Moreover, the lack of statistical influence of respondents' previous experiences with chatbots in their responses indicates that it can serve as a starting point for novices and to help improve the chatbot development for experienced developers or designers.

There were some interesting suggestions from the survey's respondents that could be implemented without losing objectivity, which is one of the guide's strengths. One of them is adding references to the SLR papers or to this paper to justify the practices and to give credibility to them. Moreover, many respondents missed a longer example of the practices. In this sense, it is possible to add another page showing a fictional chatbot that was not designed with the guidelines and another version of it improved with the proposed guidelines.

Some comments with things out of our scope indicated that many respondents did not understand the main goal. In fact, our proposed guide does not intend to teach conversational design from scratch or get into technological details, but to present practices that are beneficial to chatbot conversations. This misunderstanding can be mitigated by explaining better the guide's objective on the first pages of the guide and also establishing its limitations.

\subsection{Theoretical Contribution}

As seen in Section \ref{sec:related_works}, the interaction of conversational agents has been an object of study in several papers, but with different methodologies and focus. After presenting the results of this article, it is possible to establish how it contributes to the literature in relation to what has already been presented by these works. Table \ref{tab:related_works} shows the practices that are mentioned in similar works, although in some of them the practices are not explicitly mentioned or are part of a broader recommendation from the paper.

The results of our SLR culminated in the production of a ready-to-use guide that summarizes the results found in a way that is accessible to professionals in the field, adding a step further in relation to pure SLR studies \citep{chaves2021, sugisaki2020, rapp2021}. Moreover, these SLR studies differ from this work mainly because they focus more on the social characteristics of the chatbots and present broader discussions rather them direct guidelines. On the other hand, in our SLR we are concerned with finding software requirements ready to be implemented.  

The works that derived guidelines from user studies had a final result closer to our proposed guide \citep{amershi2019guidelines, yang2021, FEINE2019138}, however, their list is different regarding the practices that are presented because of different focus, such as considering all kinds of conversational agents or AI in general, or because of the limitation of their sample. The SLR approach enabled us to take advantage of many user studies conducted with a diverse sample, making our list of practices broader than practices elicited from single-user studies.

\begin{table}[htbp]
     \caption{Practices that were listed directly or indirectly by related works RW1\citep{chaves2021}, RW2\citep{sugisaki2020}, RW3\citep{rapp2021}, RW4\citep{amershi2019guidelines}, RW5\citep{yang2021} and RW6\citep{FEINE2019138} .}
     \label{tab:related_works}
    \scriptsize
  \centering
  \begin{tabular}{lccccccc}
    \hline
    Naturalness & RW1 & RW2 & RW3 & RW4 & RW5 & RW6 \\
    \hline
    Self-introduction   &   & \scalecheck & \scalecheck & &  & \scalecheck \\
    Address user by name &  & \scalecheck &  & &  &  \\
    Small talk & \scalecheck & \scalecheck & \scalecheck &  &  & \scalecheck \\
    Echoing responses &  &  & \scalecheck &  &  &  \\
    Casual language & \scalecheck & \scalecheck & \scalecheck & \scalecheck & \scalecheck &  \\
    \hline
    Emotionality & RW1 & RW2 & RW3 & RW4 & RW5 & RW6 \\
    \hline
    Exclamatory feedback &  &  & \scalecheck &  & &  \\
    Graphical media &  \scalecheck &  &  \scalecheck & &  &  \\
    Empathic messages & \scalecheck &  & \scalecheck & &  &  \\
    Humor &  \scalecheck &  & \scalecheck & & & \scalecheck \\
    \hline
    Transparency & RW1 & RW2 & RW3 & RW4 & RW5 & RW6 \\
    \hline
    Present capabilites & \scalecheck &  & & \scalecheck & \scalecheck &  \\
    Acknowledge limitations &   &  &  &  &  &\\
    Make suggestions &   \scalecheck & \scalecheck & \scalecheck & \scalecheck &  &  \\
    Ask for clarification &   & \scalecheck & \scalecheck &  &  \\
    \hline
    What to avoid & RW1 & RW2 & RW3 & RW4 & RW5 & RW6 \\
    \hline
    Repetitive messages &   &  &  &  &  \\
    Exaggeration &  &  &  &  &  \\
    Hiding true identity & \scalecheck & \scalecheck & \scalecheck &  &  &  \\
    Machine-like typeface &  &  &\scalecheck &  &  & \scalecheck  \\
    Forcing errors &  & \scalecheck &  \scalecheck & &  &  \\
    \hline

\end{tabular}
\end{table}

\section{Threats to Validity and Limitations}

This work suffers from some threats that are common in SLRs \citep{ZhouJZLH16}. The first threat is the use of an automatic search only, which can result in missing primary studies. Moreover, the limited number of authors may introduce a bias in the selection of studies since there are only two researchers for discussing and reaching a consensus on the inclusion or exclusion of a paper. Lastly, since the first step was excluding papers by abstract, it is possible that relevant papers were excluded due to poorly written or incomplete abstracts that do not properly convey the work that has been done.

The validation of the guide suffers from threats seen in surveys, such as the sincerity of responses and the commitment of respondents in reading the whole guide carefully, since their reading was unsupervised. The fact that participation was entirely voluntary and without incentives or gains linked to the survey response reduced the possibility of ill-considered answers. Furthermore, we added mandatory questions that made respondents confirm that they had followed the instructions before moving on to the next stages to mitigate these effects.

Concerning the limitations and coverage of this work, we only cover practices for text-based chatbots, which may be applied or adapted to speech interfaces, nevertheless, the impacts may be different from the ones presented here. The coverage of the impacts is also limited because we only considered positive outcomes in our search string since including negative keywords and their synonyms would make the string too big. Moreover, the guide only summarizes the results of the selected papers and there may be other practices and strategies that positively impact users for each purpose that were not listed. 

\section{Conclusions}

In this work, we have proposed guidelines for chatbot conversational design based on a systematic literature review (SLR) to uncover the practices that have been used in chatbot conversational design alongside their impacts on users. We selected 40 papers from different contexts and with a variety of practices being tested with users to evaluate how they feel about the presence or absence of these practices.

The joint analysis of papers revealed some patterns in chatbot design that were attached to the chatbot purpose, that is, for each purpose, papers generally focused on a group of impacts and tested practices to enhance positive impacts. These patterns were added to our conceptual map which was the starting point for the creation of a guide. 

The guide itself was developed as a web page that was shared with technology practitioners to gather their opinions about it and assess the quality of the guide. Results have shown that its main strengths are objectivity and clarity, that the guide is useful for practitioners with different levels of experience, and is generic enough for being used in various domains. 

\section*{Disclosure statement}

The authors report there are no competing interests to declare.

\section*{Funding}

This study was financed in part by the Coordenação de Aperfeiçoamento de Pessoal de Nível Superior – Brasil (CAPES) – Finance Code 001.

\section*{Data availability statement}

The data that support the findings of this study are openly available in Zenodo at \href{https://doi.org/10.5281/zenodo.6399190}{https://doi.org/10.5281/zenodo.6399190}.

\section*{Notes on contributors}

\textbf{Geovana Ramos Sousa Silva} is currently pursuing a master's degree in informatics at the University of Braslía (UnB). She is also a research assistant at the Decision Technologies Laboratory (LATITUDE). Her research interests include Virtual Assistants, Human-Computer Interaction, Software Engineering, and Programming Education.

\hspace{1pt}

\noindent\textbf{Edna Dias Canedo} received the Ph.D. degree in electrical engineering from the University of Brasília (UnB), Brazil, in 2012. She is currently an Associate Professor (tenure track) with the Computer Science Department, UnB. Her current research interests include Software Engineering, Requirements Engineering, Gender Diversity, Software Systems, Empirical Software Engineering , and Usability.

\bibliographystyle{apacite}
\bibliography{references.bib}

\bibliographystylePS{apacite}
\bibliographyPS{slr.bib}

\pagebreak
\appendix
\section{Data extracted from primary studies}\label{ap:data_extracted}

\begin{table}[htbp]
\caption{Identification of selected studies with the chatbot's context and number of participants for the user study.}
\footnotesize
\begin{tabular}{llll}
\hline
ID   & Paper                    & Context                      & Users \\
\hline
PS1  & \citepPS{Candello2017}    & Finance                      & 199   \\
PS2  & \citepPS{ARAUJO2018183}   & Shopping                     & 175   \\
PS3  & \citepPS{Fadhil2018}      & Health                       & 58    \\
PS4  & \citepPS{Gnewuch2018}     & Customer Service             & 84    \\
PS5  & \citepPS{Ashktorab2019}   & Shopping, Banking and Travel & 203   \\
PS6  & \citepPS{Diederich2019}   & Customer Service             & 112   \\
PS7  & \citepPS{Zhou2019}        & Interview                    & 1280  \\
PS8  & \citepPS{Bawa2020}        & Open-domain                  & 91    \\
PS9  & \citepPS{Beattie2020409}  & Recommendation               & 96    \\
PS10 & \citepPS{DeCicco20201213} & Delivery                     & 193   \\
PS11 & \citepPS{Diederich2020}   & Customer Service             & 77    \\
PS12 & \citepPS{Hendriks2020271} & Customer Service             & 159   \\
PS13 & \citepPS{Lee2020}         & Mental Health                & 47    \\
PS14 & \citepPS{Liao2020430}     & Mental Health                & 212   \\
PS15 & \citepPS{Narducci2020251} & Recommendation               & 54    \\
PS16 & \citepPS{Ng2020190}       & Financial                    & 410   \\
PS17 & \citepPS{SHEEHAN202014}   & Booking                      & 189   \\
PS18 & \citepPS{Shi2020}         & Donations                    & 790   \\
PS19 & \citepPS{Toader2020}      & Shopping                     & 240   \\
PS20 & \citepPS{Xiao2020}        & Interview                    & 206   \\
PS21 & \citepPS{AhmadSR21}       & Open-domain                  & 263   \\
PS22 & \citepPS{Buhrke20214456}  & Customer Service             & 228   \\
PS23 & \citepPS{Ceha2021}        & Learning                     & 58    \\
PS24 & \citepPS{DeCicco2021140}  & Delivery                     & 171  \\
PS25 & \citepPS{Kull2021840}        & Customer Service & 80  \\
PS26 & \citepPS{Ma202121}           & Shopping         & 54  \\
PS27 & \citepPS{Mozafari20212916}   & Customer Service & 257 \\
PS28 & \citepPS{Mozafari20214}      & Customer Service & 201 \\
PS29 & \citepPS{Pizzi2021878}       & Shopping         & 400 \\
PS30 & \citepPS{RanaMS21}           & Rental           & 150 \\
PS31 & \citepPS{Schanke2021736}     & Shopping         & 426 \\
PS32 & \citepPS{Spillner2021}       & Recommendation & 75  \\
PS33 & \citepPS{Tsai2021460}        & Customer Service & 155 \\
PS34 & \citepPS{Wilkinson2021}      & Recommendation & 310 \\
PS35 & \citepPS{Medeiros2021}       & Mental Health    & 210 \\
PS36 & \citepPS{Svikhnushina2021}   & Open-domain      & 536 \\
PS37 & \citepPS{Chaves2022}         & Tourism          & 178 \\
PS38 & \citepPS{Pecune2022}         & Recommendation & 289 \\
PS39 & \citepPS{EsmarkJones2022703} & Open-domain      & 139 \\
PS40 & \citepPS{Rhim2022}           & Surveys          & 59  \\
\hline
\end{tabular}
\label{tab:selected}
\end{table}

\begin{table*}[p]
     \caption{Conversational practices extracted from each selected study.}
    \scriptsize
  \centering
\rowcolors{2}{gray!10}{gray!30}
  \begin{tabular}{cp{2.5cm}p{4.2cm}p{4.2cm}p{2.5cm}}
    \hline
    ID & Strategy & Conversational practice(s) & Impact(s) on users & Moderator(s) \\
    \hline
PS1 & Typeface & Machine-like typeface (OCR-A)  & [-]perception of humanness & Familiarity with AI \\
PS2 & Anthropomorphic cues & Human name, informal language & [N]perception of social presence, [+]mindless perception of anthropomorphism, [+]mindful perception of anthropomorphism & None  \\
PS3 & Emoji based dialogue & Emoji & [+]enjoyment, [+]attitude, [+]confidence & None \\
PS4 & Social cue & Dynamically delayed responses & [+]perception of humanness, [+]perception of social presence, [+]satisfaction  & None \\
PS5 & Repair [breakdown] & Acknowledging misunderstanding and suggesting solutions & [+]preference & [User's] social orientation, experience with chatbots and technology.    \\
PS6 & Social cue & Sentiment-adaptive responses & [+]perception of empathy, [+]perception of humanness, [+]perception of social presence, [+]satisfaction &  [User's] gender \\ 
PS7 & Reserved and assertive personality & Reserved, calm, assertive, rational, careful, like a counselor & [+]willingness to confide, [+]willingness to listen & [User's] personality and context  \\
PS8 & Linguistic style & Code-mix & [+]perception of conversational ability, [+]perception of human-ness & [User's] language proficiency   \\
PS9 & Nonverbal cues & Emojis & [+]social attractiveness, [+]perception of competence, [+]perception of credibility & None \\
PS10 & Visual cues & Avatar & [N]perception of social presence & None \\
PS11 & Preset answer options & Buttons & [-]perception of humanness, [-]perception of social presence, [N]satisfaction & None \\
PS12 & Self-presentation & Introducing itself as a chatbot & [-]perception of social presence, [-]perception of humanness, [N]satisfaction  & None \\
PS13 & Self-disclosure & Small talk & [+]self-disclosure & Passage of time \\
PS14 & Racial mirroring & Profile pictures and names that might have implied their racial identity & [+]interpersonal closeness, [-]disclosure comfort, $^a$[+]satisfaction & None   \\
PS15 & Interaction modes & Buttons &  [N]satisfied, [+]understood & None \\
PS16 & Socio-emotional features & Human name, respond empathetically, give encouraging statements, active listening skills, using the user’s preferred name, turn-take and small talk & [N]trust, [N]privacy concerns, [-]data disclosure  &  Perception of social presence \\
PS17 & Repair [miscommunication] & Clarification request & [+]perception of anthropomorphism, [+]adoption intent & None  \\
PS18 & Persuasive & Inquiry & [-]donation probability & Perceived identity of the chatbot \\
PS19 & Gender cues & Female avatar & [+]forgive in the error condition, [+]satisfaction, [+]social disclosure & None \\
PS20 & Active Listening & Paraphrasing, verbalizing emotions, summarizing and encouraging & [+]engagement, [+]interest, [+]chat experience & None \\
PS21 & Extraverted & Many topics in a short amount of time, informal language, compliments and positive emotion words & [N]perception of humanness, [+]perception of social presence, [N]communication satisfaction & User personality \\
PS22 & Typing errors & Dynamic error, temporal error and spatial error & [-]perception of humanness, [-]perception of social presence & None \\
PS23 & Humor & Jokes, conundrum riddles and funny stories & [+]motivation, [+]effort & Self-defeating humour \\
PS24 & Social-oriented & Small talk, exclamatory feedback, GIFs and emoticons & [+]perception of social presence, [N]trust, [+]enjoyment, [N]intention to use & None \\
  \hline
\end{tabular}
\label{tab:practices}
  \begin{minipage}{15cm}
    \centering
    \footnotesize{
        \hfil
        [+]positive impact [-]negative impact [N]neutral impact
        \newline  \null\hfil
        $^a$ higher ``satisfaction" in comparison with ``not mirroring" but not with baseline (robot avatar);
    }
  \end{minipage}
\end{table*}

\addtocounter{table}{-1}

\begin{table*}[p]
     \caption{(continued) Conversational practices extracted from each selected study.}
    \scriptsize
  \centering
\rowcolors{2}{gray!10}{gray!30}
  \begin{tabular}{cp{2.5cm}p{4.2cm}p{4.2cm}p{2.5cm}}
    \hline
 ID & Strategy & Conversational practice(s) & Impact(s) on users & Moderator(s) \\
    \hline
PS25 & Warmth & Friendly initial message & [+]engagement & Brand affiliation \\
PS26 & Mixed-modality interaction & Buttons, sliders and checkboxes & [+]enjoyability, [N]perception of supportiveness, [N]perception of efficiency, [N]perception of precision & None  \\
PS27 & Chatbot disclosure & Introduced himself as ``Michael" and revealed himself as a chatbot & [-]trust & Acknowledge expertise or weakness  \\
PS28 & Chatbot disclosure & Introduced himself as ``Leon". [...] At the end of
the conversation, it was revealed [...] that the service agent [...] was in fact not a human person, but a chatbot & [-]trust, [-]perception of humanness & Service criticality and failure setting \\
PS29 & Conversation initiation & System-initiated non-anthropomorphic assistant & [+]reactance & Anthropomorphic avatar, [user's] gender \\
PS30 & Politeness & Polite greeting, polite goodbye, polite thanks and polite info on hours  & [+]engagement & [User's] gender, age and personality \\
PS31 & Anthropomorphism & Human name, informal language, typing cues, dynamic delay, jokes & [+]intention to buy, [+]offer sensitivity, $^b$[+]likeability & [Chatbot] disclosure \\
PS32 & Linguistic &  Lexical and structural alignment of responses & [+]user alignment & None \\
PS33 & Social presence &  Responses were designed to be informal, expressing emotions, and using numerous emojis and funny memes. Asking users their names to greet and address them by name.   & [+]user engagement, [+]satisfaction, [+]brand likeability & None \\
PS34 & Justification & Explaining why an item was recommended & [+]trust, [+]perception of transparency  & [User's] age and experience with technology    \\
PS35 & Supportive messages & Attentional deployment, cognitive change, general emotional support, situation modification & [+]valence, [N]arousal & Participants who believed to be interacting with a human being \\
PS36 & Social and Emotional Qualities & Politeness, small talk, sense of humor, emojis, short exclamations, express feelings, call me[user] by my name, ask questions.  & [+]behavioral intentions & [User's] openness to technologies, empathy propensity and vulnerability \\
PS37 & Register Compliance & Linguistic features & [+]perception of appropriateness, [+]perception of credibility & Domain \\
PS38 & Rapport-building & Justify its recommendation &[+]trust, [+]satisfaction, [+]perception of usefulness, [+]perception of ease of use & \\
PS39 & Authenticity signals & Female avatar & [+]perception of authenticity, [+]engagement, [+]satisfaction, [+]loyalty & [Avatar's] race congruence and professional dress\\
PS40 & Humanization & Self-introduction, addressing respondents by their name, adaptive response speed and echoing respondents' answers & [+]perception of anthropomorphism, [+]perception of social presence, [+]satisfaction, [+]self-disclosure & None \\
  \hline
\end{tabular}
  \begin{minipage}{15cm}
    \centering
    \footnotesize{
        \hfil
        [+]positive impact [-]negative impact [N]neutral impact
        \newline  \null\hfil
        $^b$ ``dynamic delayed response" has not improved likeability in individual experiments;
    }
  \end{minipage}
\end{table*}

\end{document}